# Log-T divergence and Insulator-to-Metal Crossover in the normal state resistivity of fluorine doped SmFeAsO$_{1-x}$F$_x$

Scott C. Riggs, J.B. Kemper, Y. Jo, Z. Stegen, L. Balicas, G.S. Boebinger
*National High Magnetic Field Laboratory, Florida State University, Tallahassee, FL, 32310*

F.F. Balakirev, Albert Migliori
*National High Magnetic Field Laboratory, Los Alamos National Laboratory, Los Alamos, NM, 87545*

H. Chen, R. H. Liu, X.H. Chen
*Hefei National Laboratory for Physical Sciences at Microscale and Department of Physics,*
*University of Science and Technology of China, Hefei, Anhui 230026, P.R. China*

We report the resistivity of a series of fluorine-doped SmFeAsO$_{1-x}$F$_x$ polycrystalline superconductors in magnetic fields up to 60T. For underdoped samples ($x < 0.15$), the low-temperature resistive state is characterized by pronounced magneto-resistance and a resistive upturn at low temperatures. The "insulating behavior" is characterized by a log-T divergence observed over a decade in temperature. In contrast, the normal state for samples with doping $x > 0.15$ display metallic behavior with little magnetoresistance, where intense magnetic fields broaden the superconducting transition rather than suppress $T_c$. The location of the insulator-to-metal crossover coincides with the reported suppression of the structural phase transition (SPT) in the phase diagram for SmFeAsO$_{1-x}$F$_x$ series.

A crucial research goal in the new oxyarsenide family of superconductors is to understand the similarities and the differences with the cuprate superconductors. For the cuprates, electrical transport measurements rapidly established key features of the phase diagram: the insulating Mott state of the undoped parent compound, the existence, amplitude and extent of the superconducting 'dome' as a function of doping, the yet-unexplained robust linear temperature dependence of the normal-state resistivity, and more recently the shape of the Fermi surface in the underdoped [1] and overdoped regimes [2].

High magnetic fields have proven valuable for studying the phase diagram of the cuprates by suppressing their superconducting phase and revealing the behavior of the normal state well below the superconducting phase transition temperature $T_c$. When Tc is suppressed by high magnetic fields, one important observation is the appearance of an insulator-to-metal crossover (IMC) in the in-plane *and* out-of-plane resistivities ($\rho_{ab}$ and $\rho_c$ respectively) for both La$_{2-x}$Sr$_x$CuO$_4$ (LSCO) [3] and Bi$_2$Sr$_{2-x}$La$_x$CuO$_{6+\delta}$ (BSLCO) [4]. These measurements revealed that the resistive upturn in this insulating regime is well-characterized by a logarithmic temperature (log-T) dependence [4,5]. The same phenomena have also been reported for the electron-doped cuprate Pr$_{2-x}$Ce$_x$CuO$_{6+\delta}$ [6].

More recent experiments used electron irradiation to induce controlled amounts of disorder into YBa$_2$Cu$_3$O$_{7-\delta}$ samples, [7,8] demonstrating that the log-T behavior is linked to the amount of disorder and scales with the square of the density of impurities. In the underdoped regime, the phenomenology of the log-T temperature dependence is consistent with Kondo scattering [8], although it has been pointed out that high magnetic fields would likely suppress conventional (spin-flip) Kondo scattering [5], suggesting instead a non-magnetic two-level scattering mechanism.

The magnetic-field-induced normal state IMC, as well as the log-T behavior, are *three-dimensional* low-temperature transport properties of the normal state regime in cuprates, as this behavior is observed independently of the orientation of the lattice with respect to the applied external field. With the goal of elucidating the *three-dimensional* properties of the low-temperature normal state in one of the oxyarsenide superconductors, we report measurements on a series of four SmFeAs(OF) samples with fluorine doping (F-doping) ranging from 0.05 to 0.20. The series of samples spans a large portion of the underdoped superconducting regime as well as optimum doping [9]. The polycrystalline samples of $SmFeAsO_{1-x}F_x$ were synthesized using conventional solid state reaction [10] and cut into rectangular prisms with a typical size of 1.5 x 1 x 0.1 mm$^3$. The resistivity ρ transverse to the applied magnetic fields was measured using the standard four-terminal digital ac lock-in technique in continous fields up to 35T and in pulsed fields up to 60T at the National High Magnetic Field Laboratory. The Tc values as measured at the midpoint of the SC transition for x=0.05, x=0.12, x=0.15, x=0.20 are ~2K, 18K, 40K, and 46K respectively.

Figure 1a shows the resistivity versus magnetic field B for our most underdoped sample, $SmFeAsO_{0.95}F_{0.05}$ at selected temperatures. Note that at 10T the magnetic field readily suppresses the superconductivity at T=0.76K, revealing the normal state resistivity at higher magnetic fields. Also note that for low temperatures (T < 20K) the normal state resistivity is *increasing* as temperature decreases ("insulating behavior"). Figure 1b contains the resistivity for our most highly doped sample, $SmFeAsO_{0.80}F_{0.20}$, in which there is no insulating behavior and the effectiveness of the high magnetic field in suppressing superconductivity is greatly reduced [11].

Figure 2 shows the evolution of the resistivity versus temperature with doping x = 0.05, 0.12, 0.15 and 0.20. Plotted are measurements from fixed-temperature magnetic field sweeps of pulsed magnets (discrete points). Dotted lines are guides to the eye connecting pulsed field data points. The most striking result is the insulating behavior of the x = 0.05 and 0.12 samples ("very underdoped"), compared to the metallic-like x=0.20 optimally doped sample.

At optimal doping, x=0.20, the superconducting state is robust under a field of 55T, which suppresses the onset of the superconducting transition by roughly 7% (Fig 2d). The midpoint of the superconducting transition is suppressed by ~20%, in effect broadening the resistive transition, as is observed in $YBa_2Cu_3O_{7-\delta}$. This is not surprising as it has been reported that the Ginzberg parameter for $SmFeAsO_{0.80}F_{0.20}$ is similar to that for YBCO [12]. Fig. 2d also shows that the magnetoresistance above $T_c$ is negligible at optimum doping: using the characteristic value of $\rho_{xx}$ ~ 1 mΩ-cm (Fig 2d) and a Hall coefficient, $R_H$ ~ -6 x $10^{-9}$ m$^3$/C [9] for $SmFeAsO_{0.80}F_{0.20}$ we estimate $\omega_c\tau$ ~ $R_H/\rho_{xx}$ ~ B[T]/1700 T, which equals 0.035 at our highest fields of 60T. Thus orbital magnetoresistance, which is of the order of $(\omega_c\tau)^2$ within a Fermi liquid picture for the $SmFeAsO_{0.80}F_{0.20}$ sample, is expected to be small in the resistive normal state.

The anomalous magnetoresistance is observed in the very underdoped samples (x < 0.15) in which the magnetoresistance exhibits a *much* larger magnitude than is observed at optimum doping even though these samples have much higher resistivity than the optimally doped sample. This is opposite to the trend expected from orbital effects. Note that the magnetoresistance extends both below and well above the zero-field $T_c$ in the very underdoped regime, persisting to temperatures as high as ~90K for x = 0.05, a temperature which is in the vicinity of the reported pseudogap [13]. It is perhaps not a coincidence that the high temperature at which the magnetoresistance becomes negligible in the underdoped regime corresponds to the same temperature where the Hall resistivity becomes non-linear in high magnetic fields [14]. The implication is that the large magnetoresistance in the underdoped regime may not be linked to the superconducting state, rather that it is a property of the normal state.

The contrast between the very underdoped samples and the optimally doped sample is most dramatic at low temperatures. At x = 0.15 (Fig 2c), the response to magnetic fields appears to be a transition between the two limiting behaviors: the magnetic field greatly broadens the resistive transition as it does at optimum doping, but it also reveals a large magnetoresistance extending well above the zero-field $T_c$. However, as is seen at optimum doping, the normal state at x = 0.15, revealed by the suppression of $T_c$, remains metallic, unlike the insulating behavior seen in the very underdoped regime. We thus conclude that the insulator to metal crossover occurs at a fluorine doping in the underdoped regime near 0.15, a doping level at which superconductivity nevertheless is quite robust ($T_c$ ~ 40K in zero magnetic field).

There is a second striking similarity between $SmFeAsO_{1-x}F_x$ and the high-Tc cuprates: the insulating behavior in the very underdoped samples can be characterized by a resistance that increases as the logarithm of the temperature. Figures 3a and 3b show the log-T resistivity of two different $SmFeAsO_{1-x}F_x$ samples with nominal x = 0.05 doping. Magnetic fields not only increase the minimum value of the resistivity, $\rho_{min}$, but also shift to higher temperatures the temperature at which the minimum resistivity occurs, $T_{min}$. Plot 3b normalizes the log-T behavior seen in the x=0.05 sample by subtracting $\rho_{min}$ and dividing temperature by $T_{min}$. Figure 3c shows the magnetic field dependence of the two parameters.

At temperatures below $T_{min}$ for $SmFeAsO_{0.95}F_{0.05}$, the resistivity appears to diverge with a log-T dependence over more than one decade in temperature, consistent with the behavior seen in the low temperature normal state properties of underdoped cuprates. For each family of materials high magnetic fields are required to reveal the log-T behavior. There are two notable differences however: (a) for the cuprates, a log-T divergence of resistivity is seen only once superconductivity is suppressed, revealing the underlying log-T normal state at low temperatures; and (b) the log-T behavior for all magnetic fields is the same. For the Sm-oxyarsenide, the onset of the insulating behavior in high magnetic fields occurs well above – as much as one decade in temperature above – the value of $T_c$ in zero magnetic field. This difference arises from the large magnetoresistance in the extremely underdoped oxyarsenides that extends to temperatures well above the log-T regime, giving rise to log-T divergences that are magnetic field dependent.

Figure 4 shows the phase diagram for temperature and doping of $SmFeAsO_{1-x}F_x$, adapted from Ref. 9. From our magneto-transport experiments (Figs 1 and 2) we shade the regions in which the large magneto-resistance and insulating behaviors are observed. Note that the SPT transition temperature [9], pseudogap energy [13] and the onset of the large magneto-resistance all occur within a relatively small window of doping from x=0.12 to 0.14. The insulator-to-metal crossover occurs at a doping very similar to the doping (x ~ 0.14) at which the SPT transition abruptly drops to zero temperature [9].

In conclusion, the resistivity of SmFeAsO$_{1-x}$F$_x$ exhibits a doping dependence with two key features in common with three-dimensional properties of the cuprate superconductors: a log-T divergence of the resistivity for more underdoped samples, and an insulator to metal crossover in the underdoped regime. A key difference between the Sm-oxyarsenide and the cuprates is the large magnetoresistance in the underdoped regime that extends to temperatures well above $T_c$ and enhances the log-T divergence of the resistivity.

Figure 1. Pulsed field measurements of resistivity versus magnetic field for fluorine doping (a) x=0.05 and (b) x=0.20 in $SmFeAsO_{1-x}F_x$. The insulating behavior can clearly be seen in the $SmFeAsO_{0.95}F_{0.05}$ sample at temperatures below 20K, while x=0.20 remains metallic at all temperatures.

Figure 2: Resistivity versus temperature of the $SmFeAsO_{1-x}F_x$ samples with four different fluorine dopings studied in magnetic fields up to 60T. Note the insulating behavior at low temperatures for samples with x < 0.15. Note also that the samples with low doping show higher magnetoresistance above Tc, despite having higher resistivities (discussed in text). The solid lines in panel (a) are logarithmic fits to the low-temperature insulating behavior.

Figure 3: Resistivity versus logarithm of temperature for two different samples of $SmFeAsO_{0.95}F_{0.05}$ from (a) magnetic field pulses up to 60T at fixed temperatures and (b) temperature sweeps in fixed magnetic fields up to 35T. The data show a weak log-T divergence of resistivity over roughly a decade in temperature. The data in (b) are scaled by $\rho_{min}$ and $T_{min}$, the resistivity and temperature at which the resistivity is a minimum. The magnetic field dependence of the normalization factors $\rho_{min}$ (right axis) and $T_{min}$ (left axis) are given in panel (c).

Figure 4: Phase diagram for $SmFeAsO_{1-x}F_x$ in a 50T magnetic field (dashed lines) from the data in Fig. 2 (filled circles), including the insulator-to-metal crossover (shaded region) and the insulating regime characterized by the log-T increase in resistivity (red). Dotted lines are the zero-field structural phase transition and superconducting transition from Ref 9.

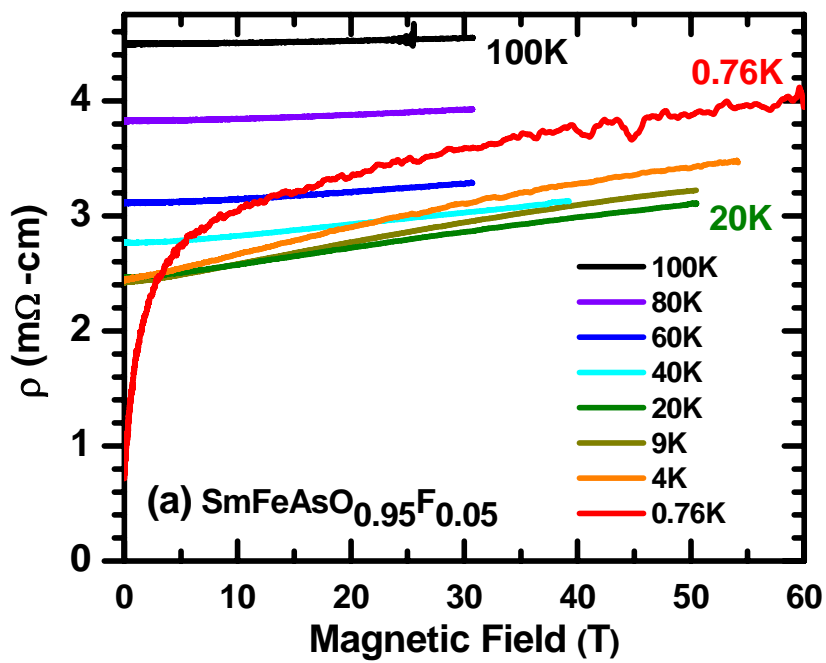
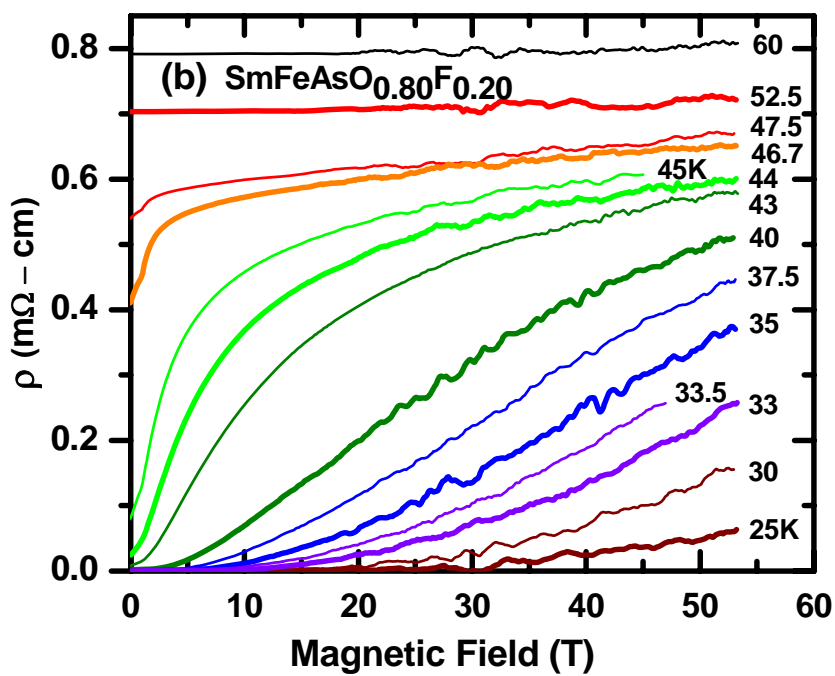

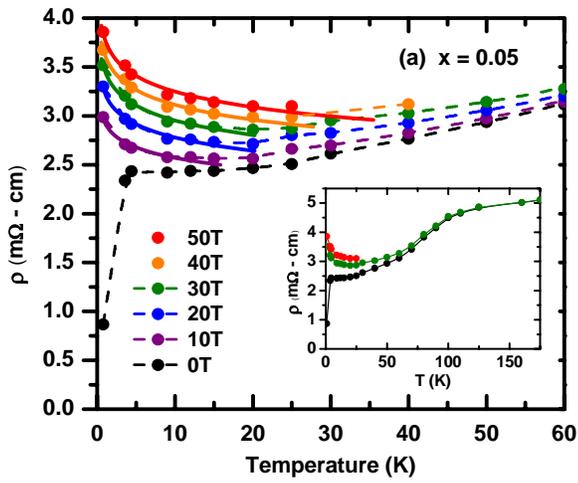
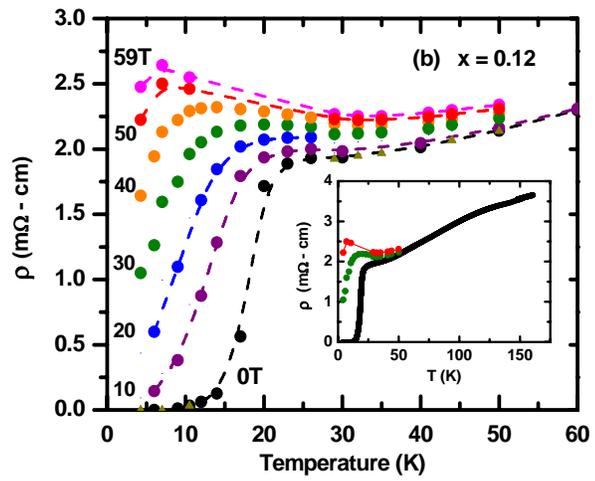
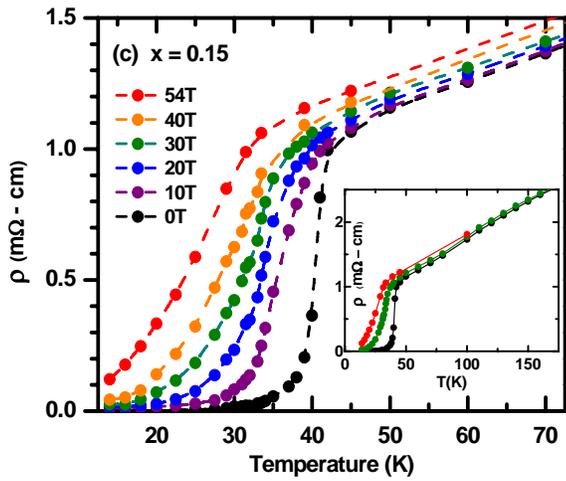
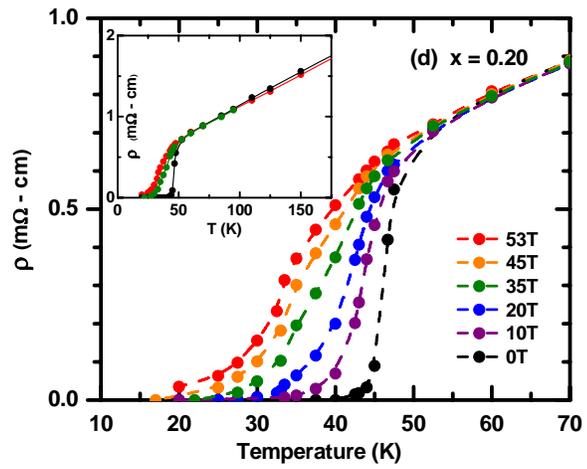

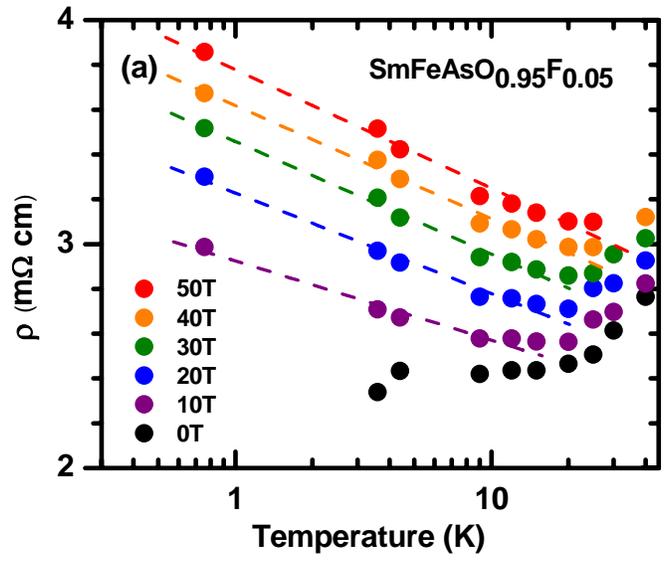
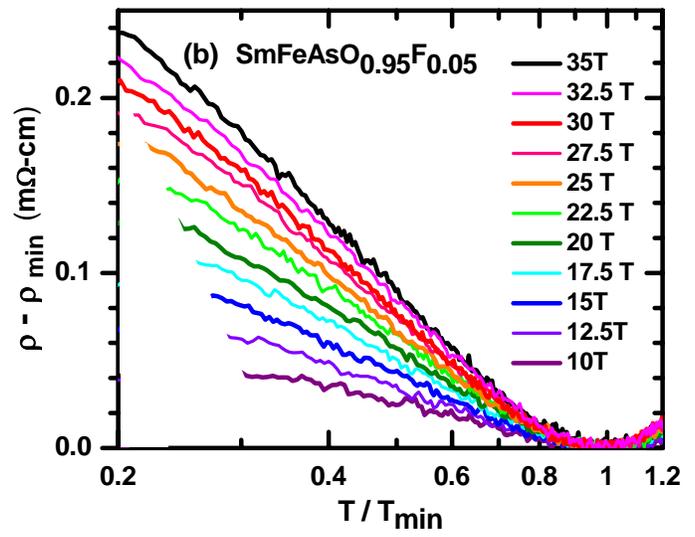
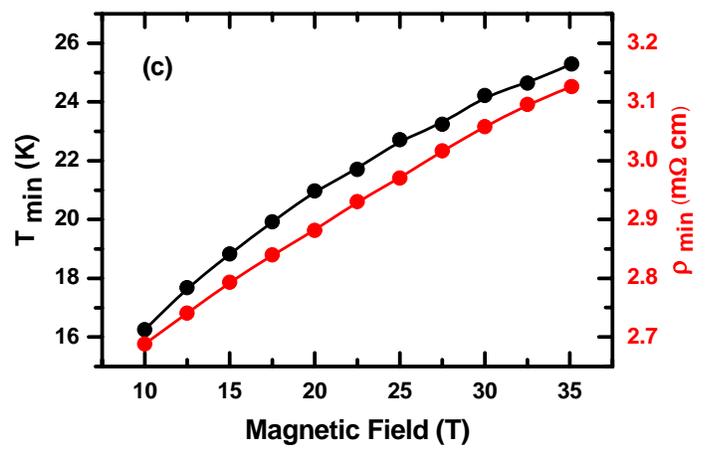

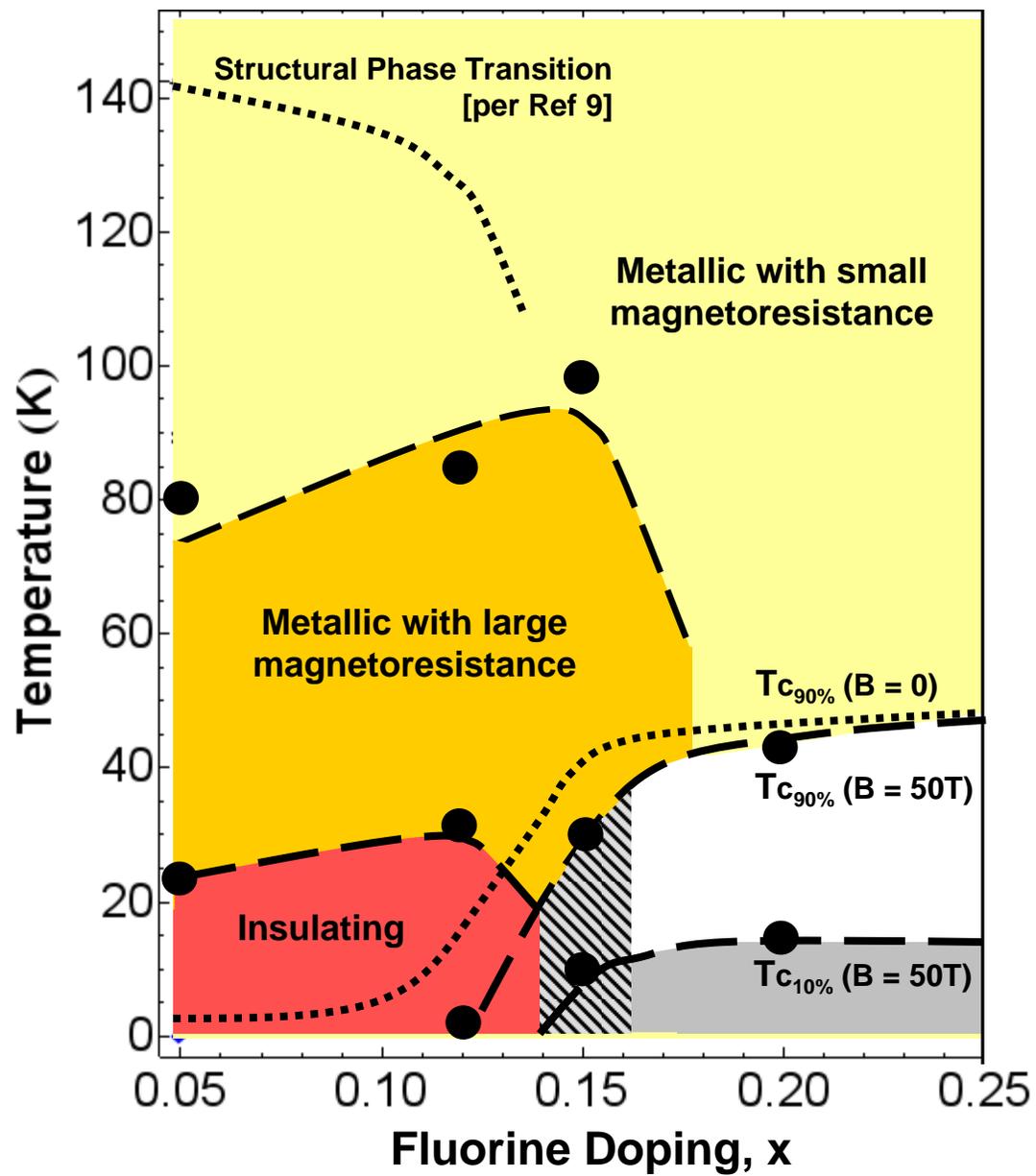